\documentclass[aps,prb,preprint,groupedaddress,epsfig,showpacs]{revtex4}
\begin{document}
\title{Trapping electrons in semiconductor air bubbles: A theoretical approach}

\author{J. Planelles}
\email{josep.planelles@exp.uji.es}
\author{J.L. Movilla}
\affiliation{Departament de Ci\`encies Experimentals, UJI, Box 224, E-12080 Castell\'o, Spain}
\date{\today}

\vspace*{1cm}

\begin{abstract}

The role of image charges in nanoporous semiconductor materials is investigated within the framework 
of the effective mass and envelope function approximations.
We show that nanometric air bubbles in these materials can act as electron-trapping centers.
This trapping capability originates from a deep stabilizing self-polarization potential well induced by the air - semiconductor dielectric mismatch which can surpass the electroaffinity barrier.
The trapping strength is a function of  the pore size and the bulk parameters of the matrix material. 
A {\it trapping parameter} characteristic for each semiconductor material is defined. This parameter provides
a simple way to ascertain the maximum pore size in a given material which is able to induce self-trapping of excess electrons.

\end{abstract}


\maketitle

\vspace{1.5cm}

The progress made in silicon technology and the increasing miniaturization connected with a simultaneously increasing integration density require the development and integration of a new generation of low dielectric constant materials.\cite{maex,petkov,bremmer,rasco,yu,treichel} The most promising strategy to diminish the effective dielectric constant of known materials relies on the introduction of pores in the material. This fact has raised vital interest in understanding the physical properties of porous materials.\cite{maex,petkov,bremmer,yu,treichel} Integration into microelectronic devices demands low pore interconnectivity and an extremely small pore size, thus forcing pores to be within the nanometric scale.\cite{maex,treichel} Pore sizes ranging from one to several tens of nanometers are nowadays commonly obtained.\cite{maex,bremmer,beji,yu}\\

Due to the different dielectric response of the involved media, non-negligible surface image charges of free electrons (or holes) may arise at the interface between the air pore and the matrix material. These charges give rise to self-polarization of the inducing carriers and polarization of the Coulomb interaction between them.\cite{rev,brus} 
The shape of the self-polarization potential is particularly sharp when the dielectric constant of one of the attached media is close to unity, yielding a narrow, deep stabilizing well near the interface that may eventually induce self-trapping of carriers.\cite{stern,proetto} 
Both experimental and theoretical evidence of carrier trapping in dielectrically mismatched surfaces can be found in the literature.\cite{biasini,goldoni_1,goldoni_2,movi_josep_wlodek,movi_josep_prb,koch_exciton} 
Since the largest effects of dielectric confinement take place in zero-dimensional nanostructures (quantum dots),\cite{QWvsQD} many works have been devoted to studying these systems under high dielectric mismatch conditions,\cite{brus,proetto,goldoni_1,goldoni_2,movi_josep_wlodek,movi_josep_prb,koch_exciton,movi_josep_cpc,tsu,iwamatsu_1,iwamatsu_2,franceschetti} particularly when the surrounding medium is air or a vacuum. However, to our knowledge, there are no studies that focus on the dielectric effects when the situation is reversed, that is, when the confined medium is air or a vacuum.\\

In this paper we study the trapping capability of conduction band electrons in spherical nanopores (air bubbles) generated in different semiconductor materials. We find that, despite the barrier-acting nature of air cavities, dielectric confinement may induce electron localization in a narrow, deep attractive well within the pore, close to its surface. Depending on the pore size and the physical properties of the semiconductor we can encounter either electron trapping or scattering by the air bubbles, which leads to opposite implications in transport through nanoporous materials.\footnote{We should point out that the studied model, a spherical isolated cavity in bulk, should be applied with caution to many current porous materials used as electron carriers in dye-sensitized and hybrid organic/inorganic solar cells and other devices\cite{devices} as many of them are porous materials with mesoscopic and nanoscopic building blocks forming connected cavities and channels with complex geometries and, often, including impurities.}\\

The present study is carried out within the framework of the effective mass and envelope function approximations. 
Consequently, we employ a macroscopic treatment of Coulombic interactions, so that a parameter, the dielectric constant, characterizes the dielectric response of each involved material. The validity of such a treatment has been well-established for zero-dimensional semiconductor heterostructures similar to those presented here.\cite{proetto}
Our model consists of a spherical air bubble ($\varepsilon_{air} = 1$, $m^*_{air} = 1$) surrounded by a semiconductor characterized by its dielectric constant $\varepsilon$ and conduction band electron effective mass $m^*$. 
The corresponding one-particle effective mass Hamiltonian reads (in atomic units),

\begin{equation}
H=-\frac{1}{2}\bigtriangledown \left(\frac{1}{m^*({\mathbf r})} \bigtriangledown \right)+ V({\mathbf r})+\phi_s({\mathbf r}),
\label{hamiltonian}
\end{equation}

\noindent where the first term is the generalized kinetic energy operator, 
and $V({\mathbf r})$ represents the spatial confining potential. When the effective-mass approach and the envelope function approximation are used, the confining potential has a well defined step-like character at interfaces separating two different media, the rectangular steps being determined by the corresponding band offsets. 
In our case, as the mismatch involves air and a semiconductor, the band offset has been assumed to be equal to the semiconductor electroaffinity. 
Finally, $\phi_s({\mathbf r})$ stands for the electron self-polarization potential coming from the image charges generated by the dielectric mismatch at the bubble edge. The calculation of this potential is carried out employing a dielectric function profile that changes smoothly within a thin interface -of the order of a lattice constant- between the air and the bulk semiconductor.\cite{proetto,movi_josep_cpc} It bypasses the (unphysical) self-polarization potential divergences that arise at the interface when a step-like dielectric profile is employed (see Refs. \onlinecite{rev,brus,stern,proetto,goldoni_1,koch_exciton,movi_josep_cpc} for details).
This model finds justification from the physical point of view, since the interface between two semiconductors (or between semiconductor and vacuum) is never perfectly sharp as the step-like model assumes.\footnote{We employ a cosinus-like dielectric profile in a 0.3 nm width interface separating the two involved media. We have checked here that, as pointed out in Ref. \onlinecite{proetto}, self-energy effects are insensitive to the smoothing model and to interface widths of the order of a lattice constant as that employed here.}
Conduction band ground state energies and wave functions were obtained by carrying out exact (numerical) integration of Eq. (\ref{hamiltonian}) by means of a discretization scheme on a grid extended far beyond the bubble radius. The explicit expression of $\phi_s({\mathbf r})$ and a detailed description of the integration method employed can be found in Refs. \onlinecite{movi_josep_prb,movi_josep_cpc}.\\

The bottom of Figure 1 shows a concrete example of the self-polarization potential profile for a $R$=5 nm air bubble generated in a $\varepsilon$=10 semiconductor material. 
As previously mentioned, the self-polarization potential yields a narrow, deep well within the nanopore, close to the interface, and has a weak repulsive character outside (see Fig. 1). The stabilizing effect of the self-energy well competes, then, with the destabilizing effect of the barrier that $V({\mathbf r})$ presents inside the nanopore, thus yielding two different situations. On the one hand, if image charge effects dominate, the electronic ground state energy arises below the conduction band edge of the semiconductor, and its wave function is mainly localized within the nanopore (see Fig. 1). We will refer to these states as $trapped$ $states$ unless an amount of the total electron density higher than 30\%  spreads outside the nanopore. On the other hand, if $V({\mathbf r})$ dominates, i.e. the semiconductor electroaffinity is large enough, the nanopore acts as a barrier, and the electronic density spreads throughout the semiconductor matrix.\\

Figure 2 represents the maximum semiconductor electroaffinity $V_{max}$ that yields trapping of electrons vs. the natural logarithm of its dielectric constant. This is shown in Fig. 2a for a fixed value of the effective mass ($m^*=0.5$) and different values of the pore radius $R$ ranging from 2 to 20 nm. It is also represented in Fig. 2b for a fixed value of the pore radius ($R=5$ nm) and different effective masses of the semiconductor matrix ranging from 0.03 up to 1, this effective mass range accounting for most semiconductor materials.\\

Several factors determine $V_{max}$. On the one hand (see Fig. 2), as the semiconductor permittivity increases, $V_{max}$ also increases because the stabilizing well of the self-polarization potential becomes deeper, and thus raises the maximum value of the barrier height that admits electron trapping.\footnote{The depth of the self-polarization potential well is related to the difference of the inverses of the permittivities involved in the dielectric mismatch, so that air-filled cavities are expected to be the most favored situation for electron trapping. Thus, an $R$=2 nm air-filled pore in a semiconductor defined by m$^*$=0.5, $\varepsilon$=10 can trap electrons overcoming a barrier height up to 2 eV. However, in the same conditions, we have calculated a barrier of just 0.6 eV if the pore is filled with a $\varepsilon$=1.78 material. Then, we cannot use Figs. 2 and 4 to determine trapping capability if the nanopore is filled with a medium other than air or a vacuum.} The bubble radius also influences $V_{max}$ (see Fig. 2a) because the self-polarization potential scales as $1/R$. Therefore, the larger the pore radius becomes, the weaker the trapping strength will be. Finally, the semiconductor effective mass $m^*$ also plays a role (see Figure 2b) because a part of the electronic density of a trapped electron leaks outside the bubble. Thus, a decrease in $m^*$ becomes an increase of the kinetic energy, i.e. an electronic energy destabilization. In other words, as $m^*$ diminishes, the trapping character of the air bubble is attenuated (see Fig. 2b). \\

An extra factor that can modify the trapping strength of a pore is a previously trapped electron. In order to analyze the magnitude of the Coulomb interaction (including polarization terms) and its influence on $V_{max}$ we have carried out CI calculations on a system of two trapped electrons employing the same scheme as in Ref. \onlinecite{movi_josep_wlodek}. The results obtained show that the total Coulomb energy contribution amounts to only about 0.1 eV, and, in consequence, the $V_{max}$ values are very slightly modified by the presence of a previously trapped electron in the air bubble. In other words, the associated Coulomb blockade effects exert a minor influence on the trapping strength of the air bubbles in comparison to the other variables $R$, $\varepsilon$, $m^*$ and $V$.\\

The lines shown in Fig. 2 represent linear regressions of the evaluated points with correlation coefficients above 0.99 in all cases. This linear behavior suggests the following exponential law relating $V_{max}$ vs. $\varepsilon$,

\begin{equation}
\frac{1}{\varepsilon}=A \exp \left(-\frac{V_{max}}{V_0}\right),
\label{exp}
\end{equation}

\noindent where $V_0$ is the slope of the lines in Figure 2 and, as we show below, only depends on $R$, thus accounting for the dependence of the trapping strength on the pore size. $A$ is a second parameter ($R$-independent) that accounts for the influence of the semiconductor effective mass on trapping. A comparison of the slopes of the (parallel) lines in Fig. 2b evidences the fact that $V_0$ is $m^*$-independent, while the change of slope vs. $R$ for a fixed $m^*$ (Fig. 2a) reveals its dependence on $R$. The common $x$-intercept of lines in Fig. 2a reveals that $A$ is $R$-independent, while Fig. 2b shows the dependence of $A$ on $m^*$. Several other representations like Fig. 2a for different $m^*$ and Fig. 2b for different $R$ (not shown) confirm the meaning and functional dependence of $V_0$ and $A$.\\

$V_0$ has a simple decreasing linear dependence on $R$ in the range of bubble radii studied (see Fig. 3) while the $m^*$-dependence of $A$ that we find in the range of masses studied (0.03 - 1) fits the following equation,

\begin{equation}
A(m^*)=0.83 \; (m^*)^{0.13}.
\label{eq3}
\end{equation}

We can define a parameter {\it t} that is characteristic of a given semiconductor bulk and which depends on the effective mass $m^*$, permittivity $\varepsilon$ and electroaffinitiy $V$ as follows:

\begin{equation}
t=\frac{Ln \left(A \varepsilon \right)}{V}, 
\label{trap}
\end{equation}

\noindent where $A=A(m^*)$ incorporates the $m^*$-dependence of  {\it t}. Equation (\ref{trap}) allows us to rewrite Eq. (\ref{exp}) as,

\begin{equation}
\frac{V_{max}}{V_0} \frac{1}{V}=t. 
\label{trap2}
\end{equation}

The trapping capability of a pore of radius $R$ in a semiconductor matrix with a given $m^*$ and $\varepsilon$ depends on the electroaffinity $V$. The condition $V=V_{max}$ defines the border between trapping and scattering. At this limit, i.e. subject to the condition $V=V_{max}$, {\it t} determines $V_0$ and therefore the radius $R_{max}$ corresponding to this border. In other words, the {\it trapping parameter t} (an intrinsic property of the semiconductor bulk) determines the largest radius $R_{max}$ that a pore can reach without losing its trapping capabilities. If the pore radius $R$ is larger than $R_{max}$ it will act as a scattering center instead.\footnote{It is worth to point out here that these results are limited to the $R$ values for which the effective mass and envelope function approximations are applicable. For extremely small bubble radii it is expected that the reduced volume of the pores would yield a large increase of the kinetic energy of the trapped electrons and, hence, a decrease of the pores trapping capacity in comparison with larger $R$ values. However, within the framework of the model employed here, this behavior holds for $R$ values lower than 1 nm in all the studied cases, where we cannot trust on the effective mass and envelope function approximations any longer (about the range of applicability of EMA see e.g. W. Jask\'olski and G.W. Bryant, Phys. Rev. B {\bf 57}, R4237 (1998); A.I. Ekimov et al., J. Opt. Soc. Am. B {\bf 10}, 100 (1993); D.J. Norris and M.G. Bawendi, Phys. Rev. B {\bf 53}, 16338 (1996)).}\\

Figure 4 presents the relation between this trapping parameter {\it t} and $R_{max}$, the maximum pore radius allowing trapping capabilities. Several specific examples of well-known semiconductor materials, namely Si, SiO$_2$, and TiO$_2$, are enclosed in the figure for the sake of illustration. The semiconductor parameters employed and the $t$ values obtained are shown in Table I. We observe that, due to the large electroaffinity of Si, pores in this material are unable to localize the electronic density, regardless of their size. However, porous SiO$_2$ admits electron trapping in pores with radii up to about 13 nm.\\

An especially interesting case is that of TiO$_2$, due to the strong influence that temperature exerts on its permittivity.\cite{parker} Dielectric constants ranging from 6 to beyond 150 have been reported in the literature.\cite{parker,mikami} 
We consider $\varepsilon=100$ and $\varepsilon=30$ as being representative of two different temperature regimes (low and high, respectively). The results obtained (Fig. 4) show that at a low temperature ($\varepsilon=100$) electron trapping holds for bubble radii up to 7 nm. This trapping capability vanishes as the temperature increases (see the case $\varepsilon=30$ in Fig. 4). We emphasize that this dependence on temperature of TiO$_2$ as a trapping/scattering center
could induce relevant electronic transport changes vs. temperature in nanoporous samples of this semiconductor.\footnote{The employed effective mass m$^*$=1 for TiO$_2$ may be found in Ref. \onlinecite{mTiO2}. However, much larger effective masses, e.g. m$^*$=10 (Ref. \onlinecite{m10TiO2}) or m$^*$=20 (Ref. \onlinecite{m20TiO2}) have been also reported. A larger m$^*$ would yield a larger value for the parameter {\it t}, i.e., a larger trapping capacity.}\\

Summing up, we have investigated the possible behavior of nanopores (air bubbles) in semiconductor materials as carrier trapping centers. We found that the dielectric confinement can be strong enough to overcome the barrier-acting nature of air, allowing the localization of electronic density within the nanopore. 
Large electron effective masses and dielectric constants, together with small electroaffinities and pore sizes, constitute the best scenario for electron trapping. A {\it trapping parameter} $t$ accounting for the contribution that the specific semiconductor matrix makes to the trapping capability of its nanopores has been defined. This trapping parameter $t$ provides a straightforward way to ascertain whether a specific nanoporous semiconductor material is able to induce electron trapping and, when this is the case, to obtain the largest radius of a trapping pore.

\begin{acknowledgments}
Financial support from MEC-DGI project CTQ2004-02315/BQU and UJI-Bancaixa project P1-B2002-01 (Spain) is gratefully acknowledged. A Spanish MECD FPU grant is also acknowledged (JLM).
\end{acknowledgments}


\newpage

\noindent Table I. Semiconductor parameters employed and $t$ values calculated (Eq. (\ref{trap})) for Si, SiO$_2$ and TiO$_2$. Two different dielectric constants are used to represent the regimes of low (T$\downarrow$) and high (T$\uparrow$) temperatures in TiO$_2$.\\

\begin{center}
\begin{tabular}{l c c c c}
\hline 
\hline 
  & $V (eV)$    & $m^*$  &   $\varepsilon$   &   $t (eV^{-1})$\\ 
\hline
Si           $\;\;$  &      4.0 $^a$      $\;\;$        &      0.26 $^b$     $\;\;$   &      12 $^a$       $\;\;$  &   0.53\\
SiO$_2$      $\;\;$  &      0.9 $^a$      $\;\;$        &       0.5 $^c$     $\;\;$   &       4 $^b$       $\;\;$  &   1.22\\
TiO$_2$      $\;\;$  &      3.9 $^a$      $\;\;$        &       1.0 $^d$     $\;\;$   &     100 (T$\downarrow$)    $\;\;$  &   1.13\\
             $\;\;$  &                    $\;\;$        &                    $\;\;$   &      30 (T$\uparrow$)    $\;\;$  &   0.82\\
\hline
\hline
\\
\multicolumn{5}{l}{$^a$ Ref. [\onlinecite{robertson}], $^b$ Ref. [\onlinecite{iwamatsu_2}], $^c$ Ref. [\onlinecite{mSiO2}], $^d$ Ref. [\onlinecite{mTiO2}].}
\end{tabular}

\end{center}

\vspace{1cm}

\begin{figure}[p]
\caption{Top: ground state radial probability density of a trapped electron in an $R$=5 nm air-filled spherical cavity generated in an m$^*$=0.5, $\varepsilon$=10 and 1.8 eV electron-affinity semiconductor material. Bottom: self-polarization potential profile ($\phi_s(r)$)  corresponding to this system. The vertical dotted line indicates the air bubble edge.}
\label{Fig1}
\end{figure}

\begin{figure}[p]
\caption{(a) Maximum semiconductor electroaffinity that allows trapping in nanopores as a function of the natural logarithm of the semiconductor dielectric constant, calculated for $m^*$=0.5 and different bubble radii. The lines displayed represent linear regressions of the calculated data. The evaluated points (19 in each series) were obtained with a precision of 0.1 eV, and have been omitted for the sake of clarity. (b) Same as (a) but for $R$ =5 nm and different values of the effective mass of the semiconductor matrix.}
\label{Fig2}
\end{figure}

\begin{figure}[p]
\caption{$V_0$ (see Eq. (\ref{exp})) vs. the air bubble radius $R$. The solid line represents the linear regression of the points displayed, with a correlation coefficient of 0.9974. The data employed corresponds to the case of $m^*$=0.5, although we essentially obtain the same fit when other $m^*$ values are used (not shown).}
\label{Fig3}
\end{figure}

\begin{figure}[p]
\caption{Maximum pore radius $R_{max}$ that allows electron trapping as a function of the trapping parameter $t$ of the semiconductor matrix. Example values for Si, SiO$_2$ and TiO$_2$ have also been enclosed.}
\label{Fig4}
\end{figure}

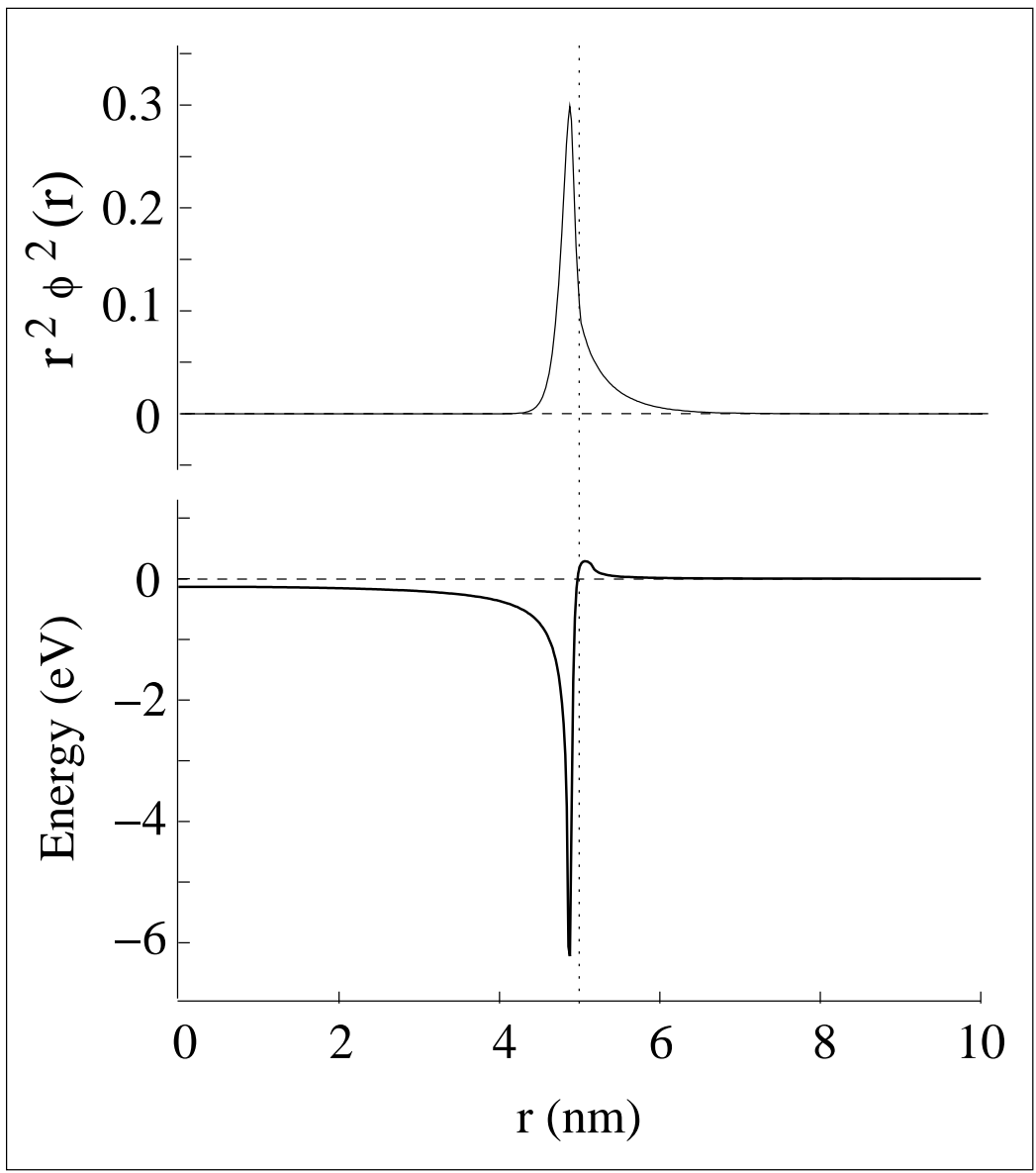
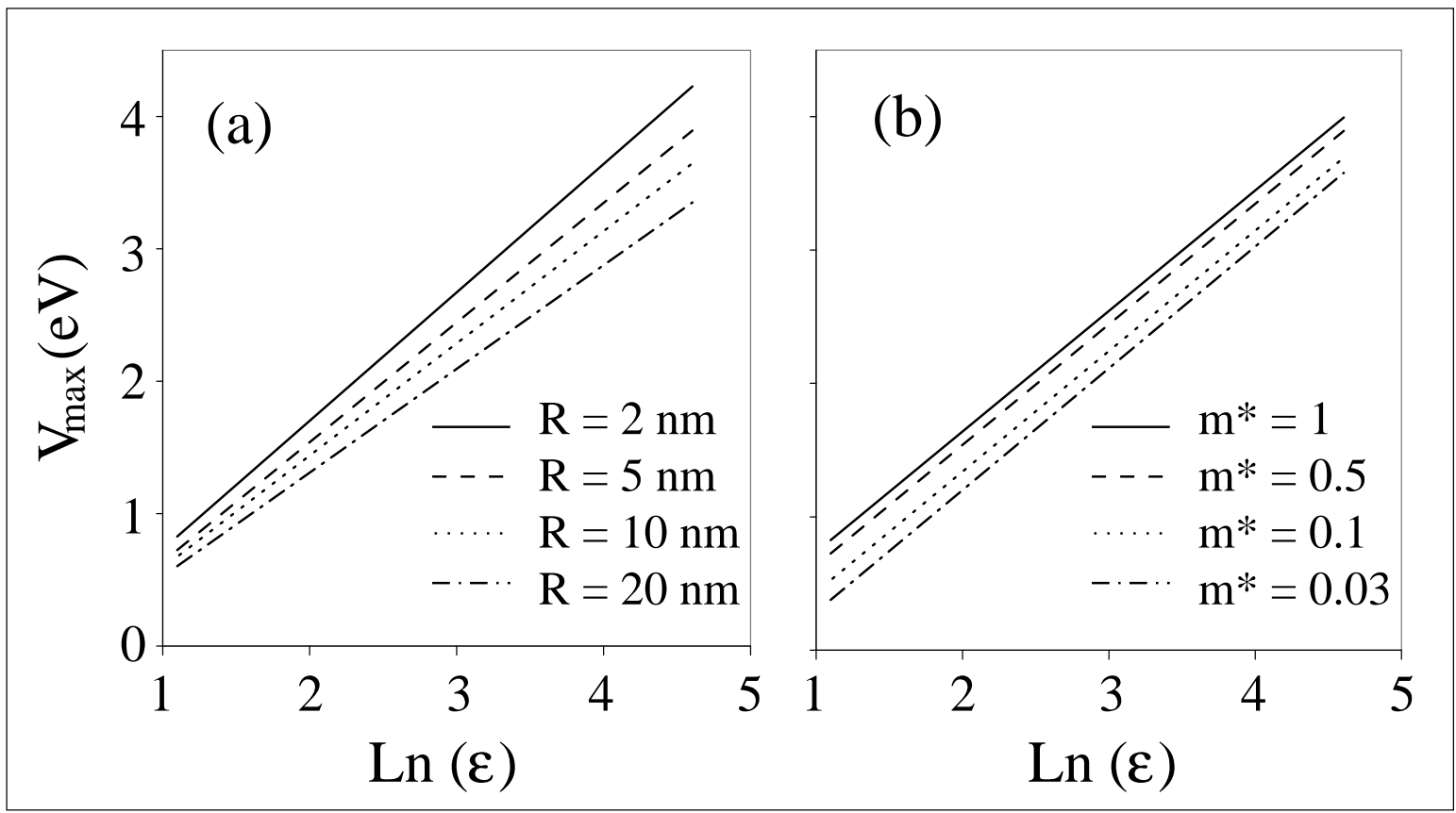
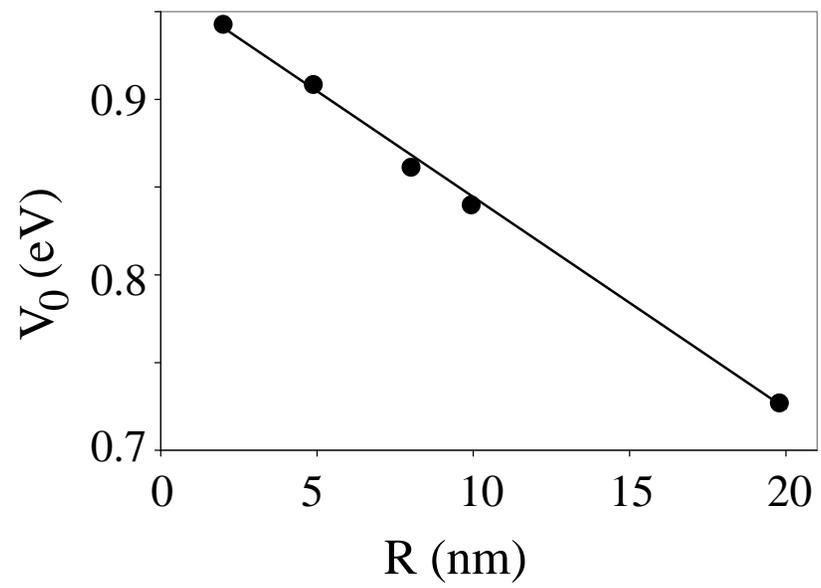
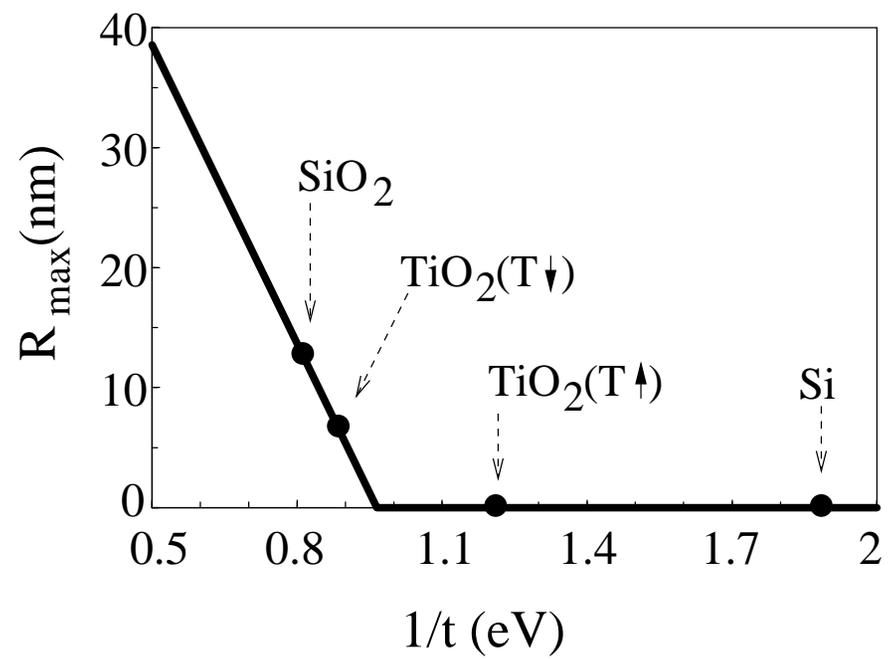
\end{document}